\documentclass[3p,times,twocolumn]{elsarticle}
\usepackage{graphicx}
\usepackage{lineno}
\journal{Nuclear Instruments and Methods A}
\begin{document}
\begin{frontmatter}

\title{Wide field-of-view Cherenkov telescope for the detection of cosmic rays\\ in coincidence with the Yakutsk extensive air shower array}

\author{A.A. Ivanov\fnref{cor}}
\fntext[cor]{Corresponding author}
\ead{ivanov@ikfia.ysn.ru}
\author{S.P. Knurenko}
\author{A.D. Krasilnikov}
\author{Z.E. Petrov}
\author{M.I. Pravdin}
\author{I.Ye. Sleptsov}
\author{L.V. Timofeev}

\address{Shafer Institute for Cosmophysical Research and Aeronomy, Yakutsk, Russia}

\begin{abstract}
The Yakutsk array group is developing a wide field-of-view Cherenkov telescope to be operated in coincidence with the surface detectors of the extensive air shower array. Currently, the engineering prototype of the reflecting telescope with the front-end electronics is designed, assembled, and tested to demonstrate the feasibility of the conceived instrument. The status and  specifications of the prototype telescope are presented, as well as the modernization program of the already existing Cherenkov light detectors subset of the array measuring ultra-high energy cosmic rays.
\end{abstract}

\begin{keyword}
Cosmic rays \sep extensive air showers \sep Cherenkov telescope \sep Yakutsk array
\end{keyword}

\end{frontmatter}

\section{Introduction}

Investigation of Cherenkov light induced by cosmic rays (CRs) cascading in the atmosphere began in the middle of the last century in the UK and the USSR. An exhaustive description of the early developments of the study of Cherenkov light is provided in \cite{Jlly}, while recent reviews of the entire area are presented, e.g., in \cite{Wtsn,Mrzn}.

It is now well-known that the angular and temporal structure of the Cherenkov light emitted by an extensive air shower (EAS) can be used to infer the longitudinal development parameters of the shower; specifically, the lateral distribution of the light intensity measured is often used to estimate the energy and mass of the primary particle initiating EAS \cite{Ztspn, Fmn, Klmkv, Trvr}. The angular distribution of Cherenkov photons from EAS was calculated by V.I. Zatsepin~\cite{Ztspn} assuming it is determined primarily by that of electrons in the shower. Subsequently, Fomin and Khristansen proposed~\cite{Fmn} to use the pulse shape of the Cherenkov signal, namely, the pulse width, to indicate the shower's maximum position, $x_m$, in the atmosphere.

Experimental measurements of the Cherenkov signal pulse shape were performed initially in Yakutsk and in Haverah Park \cite{Klmkv, Trvr}. The results were used to estimate $x_m$, and attempts were made to evaluate the cascade parameters of electrons at CR energies of approximately $10^{17}$ eV \cite{Trvr,Prsn}. A variety of detectors were used then; for example, the Tunka experiment operates an array of Cherenkov detectors near Lake Baikal \cite{Tunka2}.

Our intention to develop a Cherenkov telescope functioning as a differential detector of EAS was motivated by the possibility to measure the depth of the cascade maximum and/or the shower age via the angular and temporal distributions of the Cherenkov signal \cite{NIMA}. Combining $x_m$ and the shower age with other characteristics measured with surface detectors of the EAS array, e.g. the energy and muon content, one is able to estimate the average mass composition of CRs. Experimental arguments in elucidating the origin of the knee and ankle in the CR spectrum will be significantly strengthened by the measurements of the angular and temporal distributions of the Cherenkov signal in the energy range above $10^{15}$ eV.

Existing scenarios of CR acceleration in the sources differ in the expected model composition around the knee and in the transition region between galactic and extragalactic components \cite{Brzk}, so accurate estimation of the average mass of CR nuclei, in addition to the improved measurement of the sharpness of the knee and ankle, should allow us to discriminate some scenarios.

The paper is structured as follows. The results of Cherenkov light modeling in EAS are described in Section 2. The design and performance of the wide field-of-view (FOV) telescope prototype is described in Section 3. The first results of EAS measurements with a prototype working in coincidence with the surface detectors are presented in Section 4. Our intentions for the modernization of the Cherenkov detectors of the Yakutsk array are discussed in Section 5. A summary is provided in Section 6, followed by three Appendices.

\section{Modeling Cherenkov light induced by EAS}

\begin{figure}[t]
\center\includegraphics[width=0.95\columnwidth]{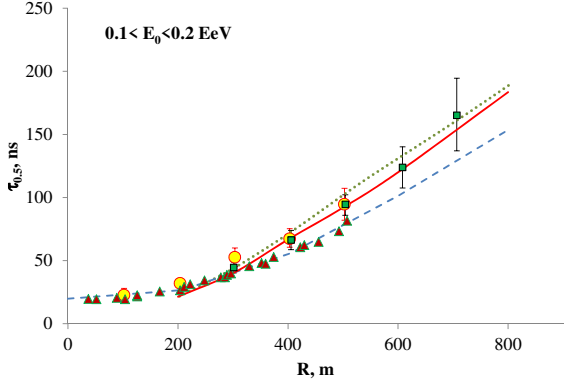}
\caption{Average pulse width at half-maximum of the Cherenkov signal vs. the shower core distance. The experimental data are from Haverah Park \cite{Trvr} (circles, $E_0=0.2$ EeV), Yakutsk \cite{Klmkv} (squares, $E_0=0.1$ EeV) and Tunka \cite{Tunka1,Tunka2} (triangles, an individual EAS event with $E_0=0.12$ EeV, $\theta=18.1^0$). Calculated widths are from \cite{Klmkv} (points), \cite{Tunka3} (dashed curve) and present work (solid curve).}\label{Fig:PulseWidth}
\end{figure}

To verify the ability of the Cherenkov telescope to discriminate the angular and temporal profiles of signals from showers initiated by primary nuclei and photons at energies above $10^{15}$ eV, we modeled the process \cite{ASTRA}. In this study, we present the results of updated simulations using a simple program described in \ref{app-a}.

The resulting full width at half-maximum of the Cherenkov signal, $\tau_{0.5}$, as a function of the core distance is compared with experimental data and previous simulations in Fig.~\ref{Fig:PulseWidth}. The approximation we used in our program is not applicable at the axis distance $R_i\leq200$ m, while the CORSIKA simulation \cite{Corsika,Tunka3} should be more precise.

An important parameter of the shower with respect to the Cherenkov light observation is the depth in the atmosphere which determines the maximum contribution to Cherenkov light observed at the particular distance from the shower axis, $x_m^{Cher}(E_0,A,R_i)$. Presumably, the depth is connected with $x_m$ at which the number of electrons reaches maximum, but is severely influenced by the angular distribution of electrons.

\begin{figure}[t]
\center\includegraphics[width=\columnwidth]{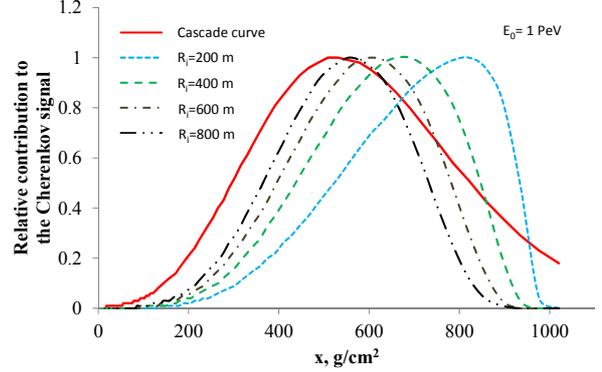}
\caption{Calculated contributions of the different depths in the atmosphere, $x$, to the Cherenkov light intensity detected at the shower axis distance, $R_i$.}\label{Fig:Xcontribution}
\end{figure}

Fig.~\ref{Fig:Xcontribution} shows the relative contributions of the depth intervals in the atmosphere to the Cherenkov light flux detected at $R_i$. For reference, the original cascade curve \cite{Knpp} is also shown. Obviously, only at large distances $R_i>800$ m does $x_m^{Cher}$ converge to $x_m$.

It was demonstrated previously, particularly in \cite{Ztspn, Trvr}, that the temporal and angular characteristics of the Cherenkov signal are related to the shower core distance and $x_m$. We confirm the core distance dependence and interpret $x_m$ dependence as a particular case of the more general shower age dependence. In this study, we define the age of the shower at the array as $\hat{s}=x_0\sec\theta/x_m$, which is connected with the cascade theory definition: $\hat{s}=2s/(3-s)$. As a result, the zenith angle dependence of the signal in the model is parameterized by the shower age.

\begin{table}[b]\centering\small
\caption{Measurable temporal and angular characteristic rates of the Cherenkov light from EAS. Notations: $t_{max}$ is the time of the Cherenkov signal maximum; $\Delta\nu$ is the width of the angular distribution of the Cherenkov signal; $\nu_{max}$ is the angle between the zenith and the direction to the maximum of light.}
\begin{tabular}{rrrr}
\hline
                          Rate & Present work & \cite{Trvr} & \cite{Ztspn}\\
\hline
    $d\tau_{0.5}/d\hat{s}$, ns &$-101.0\pm30.0$&           &    \\
       $dt_{max}/d\hat{s}$, ns & $-58.0\pm15.0$&           &    \\
  $d\tau_{0.5}/dR_i$, ns/100 m & $ 29.4\pm 3.0$&$24.0\pm3.0$&    \\
     $dt_{max}/dR_i$, ns/100 m & $ 29.7\pm 4.0$&           &    \\
    $d\Delta\nu/d\hat{s}$, deg & $-1.39\pm0.20$&           &    \\
    $d\nu_{max}/d\hat{s}$, deg & $-2.31\pm0.30$&           &    \\
  $d\Delta\nu/dR_i$, deg/100 m & $ 0.24\pm0.08$&           &    \\
  $d\nu_{max}/dR_i$, deg/100 m & $ 0.76\pm0.11$&           & $0.5$\\
\hline
\end{tabular}
\label{Table:1}\end{table}

Our simulation results are presented in Table \ref{Table:1} together with appropriate previous results. Assuming in accordance with calculations in \cite{Rss} of the age difference of showers initiated by the primary proton and iron nucleus of energy of 10 PeV, $\Delta\hat{s}=0.38$, and proton and photon initiated showers, $\Delta\hat{s}=0.15$, we conclude that the resolution of better than $15\pm 5$ ns in $\tau_{0.5}$ and $9\pm 2$ ns in the time of the signal maximum, $t_{max}$, is sufficient to distinguish the primary particles.

Similarly, the angular resolution should be better than $0.2^0$ in the width of the angular distribution and $0.4^0$ in the direction to the maximum of intensity.

\begin{figure}[t]\centering
\includegraphics[width=\columnwidth]{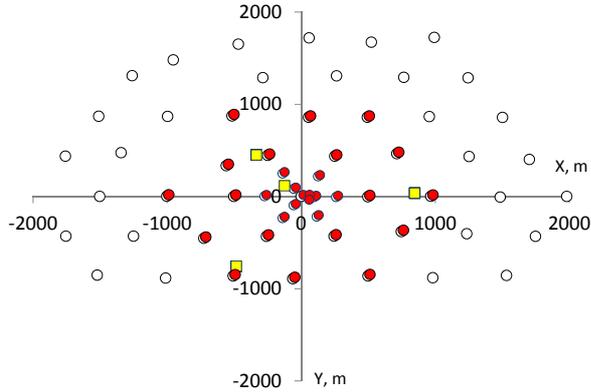}
\caption{Plan of the Yakutsk array. Open circles indicate soft component detectors, while solid circles indicate Cherenkov light detectors and squares indicate muon detectors.}
\label{Fig:Map}\end{figure}

\section{Technical design of the Cherenkov telescope}

Currently, the Yakutsk array is measuring the soft component of EAS with 58 plastic scintillators, muons with 4 underground detectors, and Cherenkov light with 48 PMTs. All the detectors are irregularly distributed within the 10 km$^2$ array area (Fig. \ref{Fig:Map}); the target energy range of investigations is $10^{15}$ eV to $10^{19}$ eV. More experimental details and physics results are given in \cite{site,MSU,APJ}.

Our plan for the future is to modernize the array to have a precise instrument capable of measuring the highest energy galactic CRs - their sources, energy spectrum, and mass composition. Another interesting object to study is a transition region between the galactic and extragalactic components of CRs, where some irregularities in the spectrum and composition may be revealed. In particular, an open question is whether the changes in the energy spectrum index at the knee and/or ankle are related to the transition between components. A possible solution may be in the different composition of CRs from galactic and extragalactic sources manifested in the transition region.

A crucial clue to this mission is the accurate determination of the mass composition of CRs, which is a weak point of the existing EAS arrays. In this context, we must adapt the well-known atmospheric Cherenkov telescope technique \cite{IACT} to measure the angular and temporal structure of the signal connected to the longitudinal shower profile above $E_0=10^{15}$ eV.

The idea is not to concentrate on discriminating the hadronic shower vs. the gamma-ray initiated shower; instead, all showers from different primary particles are detected by the wide field-of-view (WFOV) telescopes in coincidence with the array detectors. Subsequently, the angular and temporal parameters of the showers measured should be analyzed to identify EAS initial particles.

\begin{figure*}[t]
\center\includegraphics[clip,width=0.95\textwidth]{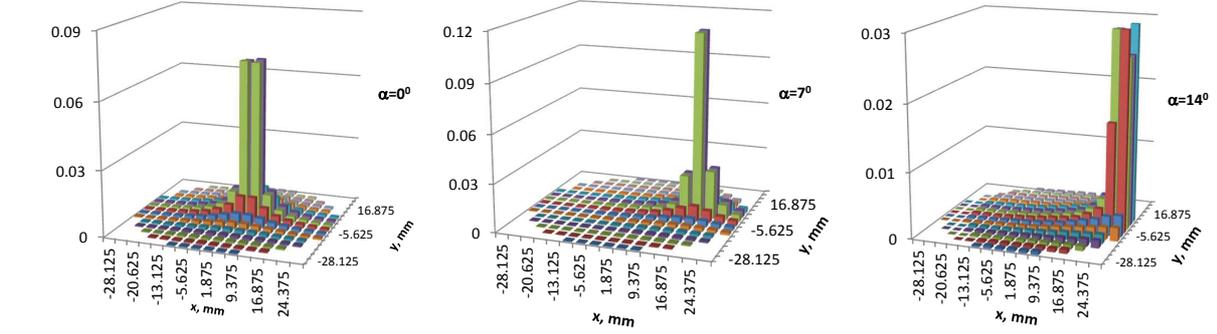}
\caption{Simulated (Monte Carlo ray tracing) image of a distant point source on the photocathode surface. The light intensity is given in arbitrary units in three cases, where $\alpha$ is the angle between the line to source and the optical axis. The pixel size is 3.75 mm$\times$3.75 mm.}\label{Fig:Image}
\end{figure*}

\begin{figure}[t]
\center\includegraphics[width=0.74\columnwidth]{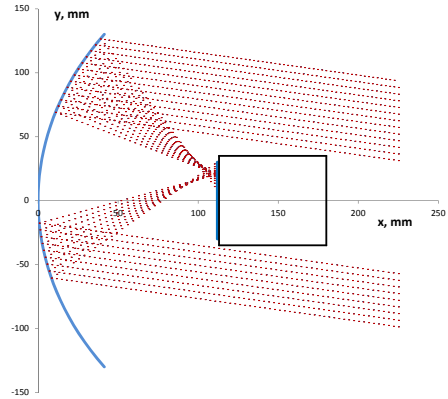}
\caption{Modeling rays (red points) from a distant point source in the telescope. Blue curves illustrate the mirror and photocathode surface. The black rectangle imitates shadowing by the PMT case.}\label{Fig:Rays}
\end{figure}

Another important step in the modernization is the essential reduction of the size and cost of the telescope. It is possible to accomplish this step due to the increased threshold energy because the total number of Cherenkov photons emitted by EAS is proportional to the shower energy. If one compares one of the HESS telescopes (diameter $D=13$ m, $E_{thr}=10^{11}$ eV \cite{Hess}) with that reduced to $D=13$ cm, then the number of Cherenkov photons from EAS detected is comparable at energy $E_0>10^{15}$ eV. Hence, in our energy range, the set of Cherenkov telescopes with $D=13$ cm is approximately equivalent to HESS functioning in the energy range $E_0>10^{11}$ eV; of course, except for the event rate: the CR intensity ratio in two energy intervals is $J(E_0>10^{15})/J(E_0>10^{11})=10^{-4\kappa}=1.6\times10^{-7}$, where $\kappa\simeq1.7$ is the  index of the integral cosmic ray spectrum.

\subsection{Ray tracing in a Newtonian telescope}

One of the possible designs of the WFOV Cherenkov telescope consists of a spherical mirror and a multi-anode PMT as an imaging camera in the focal plane. To model the focusing of the aluminized spherical mirror in the wavelength interval (300,600) nm, we used a point source of light placed at infinity, with angle $\alpha$ between the line to the source and the optical axis of the mirror. The image of the point source is calculated on the target plane near the focus of the mirror. The mirror size, $D_{mirror}$, radius of curvature, $R_{mirror}$, and target position, $F$, must be optimized to obtain as wide an FOV as possible, where the size of the blurred image is comparable to the pixel size of the position-sensitive PMT.

We used a Hamamatsu R2486 series PMT with a $16\times16$ crossed wire anode as an imaging camera of the prototype telescope. With an effective area determined by $D_{PMT}=50$ mm and the distance between wires $d=3.75$ mm providing a pixel size of approximately $d\times d$, we determined the optimal parameters of the telescope to be: $D_{mirror}=260$ mm; $R_{mirror}=225$ mm; $F=113$ mm.

Ray tracing in geometrical optics is based on the rectilinear propagation of light and reflection from the mirror surface. For a distant point source, an added bonus is the parallelism of rays. Therefore, we must solve a system of two equations to find the reflection point: 1) ray trajectory, $\overrightarrow{r}=\overrightarrow{r_{0}}+\overrightarrow{v}t$, where $\overrightarrow{r_{0}}$ is initial coordinate; $\overrightarrow{v}$ is light velocity, and 2) the spherical mirror surface, $|\overrightarrow{r}-\overrightarrow{r_c}|^2=R^2$, with radius $R$ and center $\overrightarrow{r_c}$.

The resultant 3D image of the intensity distribution of light in the target plane is presented in Fig. \ref{Fig:Image} for three typical incident angles. Fig. \ref{Fig:Rays} illustrates ray tracing in the plane crossing the optical axis; the spherical aberration of the point source image is seen on the PMT photocathode surface. The simulated FOV of our prototype telescope is found to be $28^0$.

\subsection{Position-sensitive PMT as an imaging camera}

There are two types of reading out the signal from position-sensitive PMTs: multi-anode and crossed wire anode. The advantages of the former are the uniformity of the output signals and the unique determination of the location on the anode area. Crossed wire anode signals are less uniform, and the location of the signal is not unambiguous in the case of a smeared signal. Nevertheless, we chose to use a Hamamatsu R2486 PMT with a crossed wire anode for our telescope due to the low price and small number of output channels required.

\begin{figure}[t]
\includegraphics[width=\columnwidth]{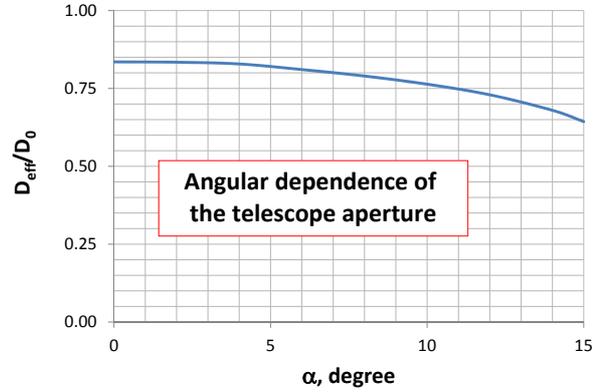}
\caption{Shadowing of the mirror by the PMT and support + cables. Plot of the ratio of the effective mirror diameter to the unshielded diameter as a function of incidence angle, $\alpha$.}
\label{Fig:Aperture}\end{figure}

Indeed, in this case, our data acquisition system (DAS) has 32 independent channels instead of 256 for the multi-anode alternative. To reconstruct the signal distribution over the anode surface, one has a system of 32 equations
\begin{eqnarray}
\sum_{j=1}^{16}{V_{ij}}=q^x_i, i=1,..,16,\\
\sum_{i=1}^{16}{V_{ij}}=q^y_j, j=1,..,16,
\end{eqnarray}
where $V_{ij}$ is the unknown signal in a grid knot; $q^x_i, q^y_j$ are the output signals of wires. The equations are dependent because of $\sum{q^x_i}=\sum{q^y_i}$. If the light source induces signals in more than one $x-$ and $y-$ wire, then there is an infinite number of solutions of the system. Only in the case of a symmetric signal with a single maximum is there a possibility to locate the position of the maximum on the anode surface.

We composed a particular solution of the system as:
\begin{equation}
V_{ij}=q^x_iq^y_j/(\sum{q^x_i}\sum{q^y_j})
\end{equation}
to locate the position of the signal maximum on the anode surface.

In the design chosen, the telescope provides an effective aperture $D_{eff}(0^0)=21.8$ cm due to shadowing of the mirror by the PMT and the support. The angular dependence of the telescope aperture is given in Fig. \ref{Fig:Aperture}. We calculated the angular dependence as a ratio of the light intensity on the photocathode surface to the initial intensity falling into the actual aperture of the telescope, taking into account the reflectance of aluminum, 92.4\%, in the PMT sensitivity interval $\lambda\in(300,600)$ nm.

\begin{figure}[t]
\includegraphics[width=\columnwidth]{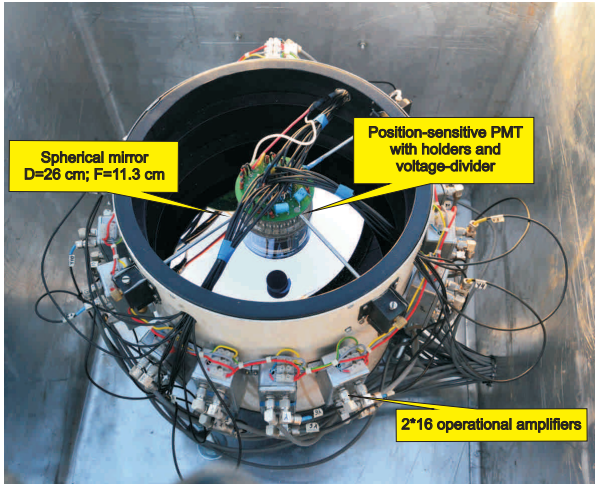}
\caption{Photograph of the WFOV Cherenkov telescope.}
\label{Fig:Telescope}\end{figure}

\begin{figure}[t]
\center\includegraphics[width=0.85\columnwidth]{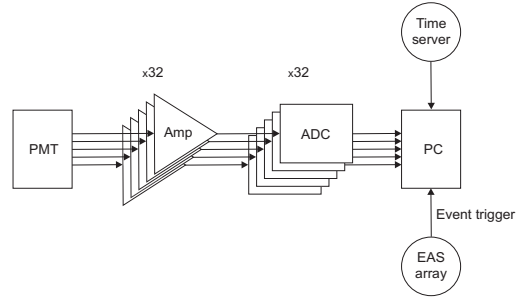}
\caption{Block diagram of the signal processing.}
\label{Fig:BlockDiagram}\end{figure}

\subsection{A telescope mount}
The telescope is fixed vertically staring at the zenith. A set of adjustment screws for the mirror and PMT positioning is used to focus the shower image.

The telescope is shown in Fig. \ref{Fig:Telescope}. A spherical mirror is mounted at the bottom of the telescope housing with vertical adjusting bolts beneath. The support sticks of the PMT are additionally used to position the PMT exactly in the focal plane.

A voltage-divider circuit and 32 signal cables are attached to the bearing plate. 16 two-channel operational amplifiers are mounted onto the telescope housing. A block diagram illustration of the DAS is shown in Fig. \ref{Fig:BlockDiagram}.

\begin{figure}[t]
\center\includegraphics[width=0.8\columnwidth]{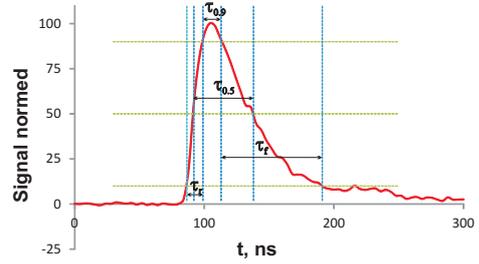}
\caption{A signal from a plastic scintillator. Pulse shape parameters are denoted.}
\label{Fig:Scint}\end{figure}

\subsection{DAS and synchronization of EAS events}
The combination of a current-to-voltage conversion circuit using operational amplifiers and ADCs connected by long (12 m) coaxial cables enabled the outdoor measurement of output signals of a PMT. On-line transfer of digital signals from 32 independent channels to a PC via USB connectors provides oscillograms of the Cherenkov light signals.

The amplifiers based on 300 MHz bandwidth AD8055 chips are used to convert the currents. As a digitizer, we used 8-bit LA-n4USB ADC with a 250 MHz sampling rate to reduce the total cost of the DAS.

The cross-talk output signals, $U_{ct}$, from the adjacent channels when the input signal, $U_{in}$, is present in a single channel were measured. The resultant ratio $U_{ct}/U_{in}$ is found to be less than 0.01 for all 31 channels, within the entire interval of the possible $U_{in}$ values.

A synchronization of the signals from the DAS with EAS events is performed as follows. All the ADC output signals from 32 channels are continuously stored in a PC memory. A trigger signal from the EAS array terminates the process, and signals in a 16 $\mu$s interval preceding a trigger are dumped. After that, the system is ready to detect the next EAS event.

Alternatively, autonomous detection of EAS events by the system is available with subsequent synchronization to the array trigger using the time stamp of the event. In this case, the system is self-triggered by a signal from the channel specified.

To test the DAS response and to illustrate the pulse shape parameters, we measured the signal from the PMT of the WFOV telescope (Fig. \ref{Fig:Scint}) induced by a plastic scintillator ($50\times50\times5$ cm) placed on the telescope inlet under the light-proof lid. The scintillator is composed of 2\% $p$-terphenyl, 0.02\% C$_{24}$H$_{16}$N$_2$O (POPOP) suspended in polymethylmethacrylate.

The normalized signals from 32 channels are averaged in the figure, and the pulse shape parameters measured are presented in Table \ref{Table:2}. In addition, the mean value and the RMS deviation of the pulse are presented in the table.

The parameters measured concern the pulse width and the rise/fall time of the signal:
\begin{itemize}
\item The rise time ($0.1A-0.9A$), $\tau_r$, where $A$ is the maximum amplitude of the signal,
\item The full width at half-maximum ($0.5A-0.5A$), $\tau_{0.5}$,
\item The top width ($0.9A-0.9A$), $\tau_{0.9}$,
\item The fall time ($0.9A-0.1A$), $\tau_f$.
\end{itemize}

The full width at half-maximum (FWHM) of the scintillator signal is $\tau_{sc}\sim5$ ns, while the DAS output signal is $\tau_{0.5}=45.4\pm8.9$ ns.

\begin{table}[t]\centering
\caption{Pulse shape parameters of the signal from a plastic scintillator measured using the telescope.}
\begin{tabular}{rrr}
\hline
          Parameter & Mean value &  RMS\\
\hline
     $\tau_{r}$, ns &       12.3 &  3.5 \\
   $\tau_{0.5}$, ns &       45.4 &  8.9 \\
   $\tau_{0.9}$, ns &       15.2 &  5.9 \\
     $\tau_{f}$, ns &       70.7 & 25.0 \\
  Pulse moments, ns &       28.1 & 22.0 \\
\hline
\end{tabular}
\label{Table:2}\end{table}

\begin{figure*}[t]\centering
\includegraphics[width=0.9\columnwidth]{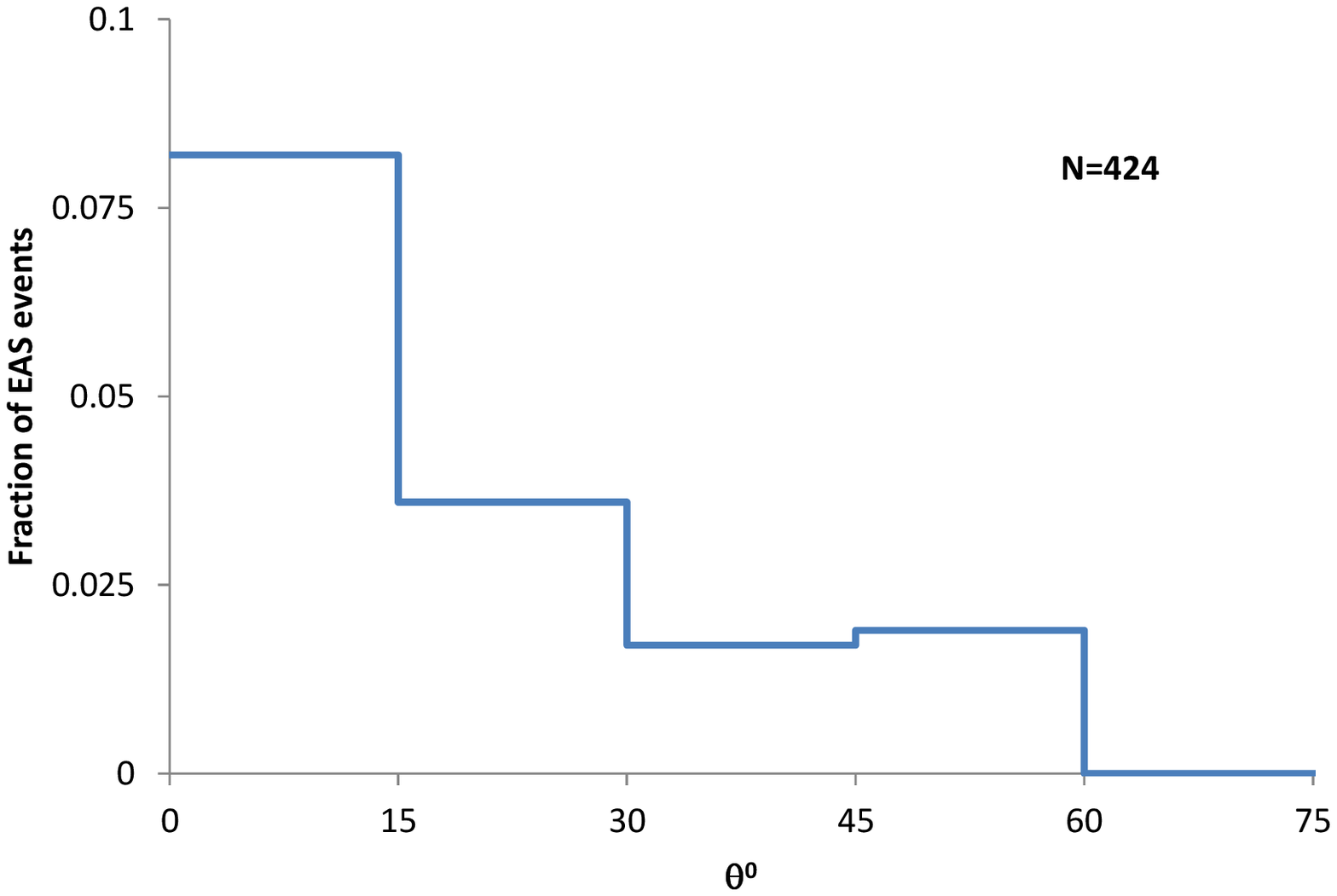}
\includegraphics[width=\columnwidth]{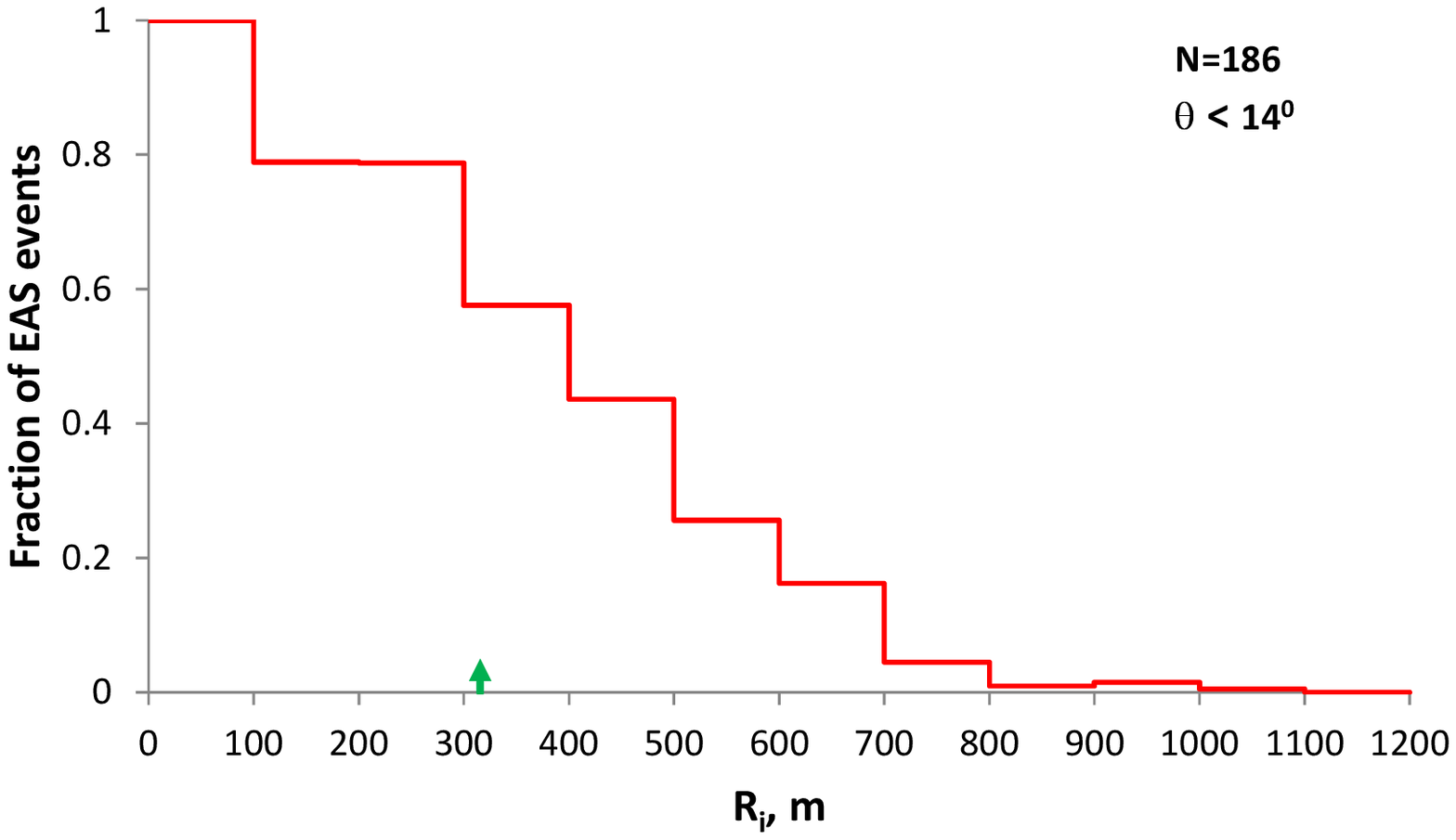}
  \caption{Zenith angle and axis distance distribution of the fraction of triggering EAS events with a detected signal in the telescope.}
  \label{Fig:Hist}
\end{figure*}

The DAS output signal is a transform of the initial scintillator signal to the response function of the data recorder. Here, the approach described in \cite{Grigoriev} is used to consider the impulse response of our system. A basic concept is the time constant of the DAS caused by the delta-function input signal. In this case, any output signal is given by a convolution of the input, $f_{in}(t)$, and the response functions, $g(t)$ (\ref{app-c}).

In a simple approximation of the similarity of signals, the DAS time constant can be estimated as $\tau_{DAS}=\sqrt{\tau_{0.5}^2-\tau_{sc}^2}=45.1\pm9$ ns, and the standard deviation $\sigma_{DAS}=\sqrt{RMS^2-\sigma_{sc}^2}=21.9\pm5$ ns.
The scintillator is used only to test the prototype; the regular telescope will be calibrated in another way.

\subsubsection{Equalizing channels with the night sky light background}
There are constant shifts in the average zero signals of channels inherent to the readout equipment. We measured signals\footnote{triggered by EAS array} from the telescope with a light-proof lid covering the telescope to compensate for the zero shifts. Based on monthly measurements in 4-hour sessions, the average zero shifts of 32 channels were found. RMS deviation of zeros in sessions is 0.5\%, and monthly deviation of zeros is less than 3\%.

Due to the unequal length of the PMT wires, telescope construction features etc., the signals of the wires are not equal, even from the same light source at different angles. To take into account the angular dependent correction to signals of channels, we have used the night sky light background (NSLB).

In moonless nights around the zenith the nearly isotropic NSLB is mainly composed of the air glow in the upper atmosphere induced by photo-chemical processes ($\sim2/3$), zodiacal light ($\sim1/3$), integrated star light, etc. The zenith angle dependence of the light intensity is approximated by $I(\theta)=I(0)(1+\theta^2/2)$, with $\theta$ in radians \cite{NSLB}. Within the FOV of our telescope mounted vertically, the background variation is less than $3\%$.

The NSLB signal from the wire at distance $x$ from the center of the photo-anode with radius $R$ is $I(x)\simeq I_0(1+\frac{R^2}{6F^2}+\frac{x^2}{3F^2})$, where $I_0$ is a signal from the isotropic NSLB; $F$ is the focal length of the mirror.

We measured NSLB reference signals in 4 ns bins with a gate duration of 16 $\mu$s during a moonless night on 07.12.2012, excluding twilight hours. The averages of the overnight NSLB signals are then used as units in measuring Cherenkov light signals from EAS in 32 channels. This approach provides equalized signals of channels but the dependence of the absolute value on the Cherenkov light intensity is still unknown.

Typical signal-to-noise ratio\footnote{treating NSLB as a noise} varies within the (2,100) interval. We have used in the analysis EAS events where this ratio is larger than 10, except for a sample of `nonzero' signal events where the ratio is larger than 2 in at least 10 independent channels.

\subsubsection{Calibration of the integral telescope signal}

We mounted the telescope near (2 m apart) one of the Yakutsk array Cherenkov light detectors. The signal of this detector was calibrated using a plastic optical radiator \cite{JETP,NJP}. Thus, we can normalize an integral signal from the telescope to that of the Cherenkov detector.

A disadvantage of this task is that the present Cherenkov detector has a comparatively large time constant and a wider FOV than our telescope. It is perhaps preferable to use another `integral' detector with the same temporal resolution and FOV.

The only device similar to our WFOV telescope in the target energy range is worked up in the Large High Altitude Air Shower Observatory (LHAASO) project in Tibet \cite{LHAASO}. Here, two prototype telescopes\footnote{with a field of view $14^0\times16^0$} are deployed near the ARGO-YBJ carpet detector array, having a 4.7 m$^2$ spherical mirror with an imaging camera consisting of $16\times16$ PMTs. The front end electronics are based on 50-MHz flash ADCs. Our prototype WFOV telescope is much more compact and cheap; DAS based on the fast electronic components allows precise waveform measurement of Cherenkov light signals from EAS.

\section{First results of EAS measurements with the Cherenkov telescope operating in coincidence with the Yakutsk array}
During the field testing of the telescope in the period from 19.10.2012 to 11.04.2013 we had 604 hours of clear moonless nights that yielded 11124 EAS events detected by the scintillator subset of the array, from which 424 events resulted in a nonzero simultaneous signal in the telescope. An additional 277 telescope signals were triggered by the Cherenkov detectors; these will be analyzed later, and the results published in a forthcoming paper.

\begin{figure*}[t]\centering
\includegraphics[width=\columnwidth]{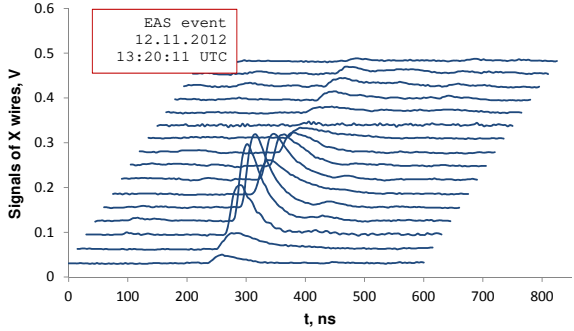}
\includegraphics[width=\columnwidth]{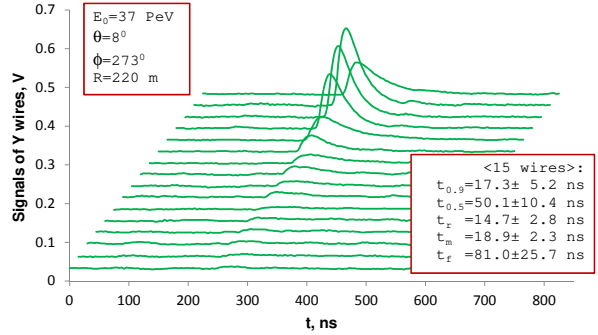}
  \caption{The Cherenkov pulse shape detected using 32 channels of the telescope DAS: 16 X wires - left panel; 16 Y wires - right panel.}
  \label{Fig:Pulses}
\end{figure*}

From the numbers above (and the array area of 8.2 km$^2$) we determined the effective radius of the telescope acceptance $R_{eff}=315$ m. This radius can be used in planning of the grid of telescopes. The fraction of EAS events detected with a nonzero telescope signal is $3.8\%$. The dependence of the signal on the zenith angle and the shower axis distance is illustrated in Fig. \ref{Fig:Hist}; the primary energies were in the region $E_0>10^{16}$ eV.

While the telescope FOV is 308 sq. degrees ($\theta\in(0^0,14^0)$), EAS events were detected with zenith angles up to $\theta=60^0$. This observation can be explained by the broad angular distribution of emitting electrons in the shower and the photon scattering in the atmosphere. Another contribution is the angular uncertainty of the shower reconstruction procedure, which is considerably increased at the lower threshold energy of the array.

To avoid this uncertainty, we selected showers within $\theta<14^0$ to define a distribution of the shower axes (right panel of Fig. \ref{Fig:Hist}). The effective radius of the telescope detecting area, $R_{eff}$, is indicated by the arrow on the $R_i$ axis as well.

We have measured the pulse shape parameters of the Cherenkov signal in individual showers. A typical set of signals from the EAS event detected on 12.11.2012, 13:20:11 UTC is shown in Fig. \ref{Fig:Pulses}. A large part of the 16 $\mu$s buffer data is cut out where only noise is recorded, so the point $t=0$ is arbitrary in this case. The signals of channels are not equalized here. The pulse shape parameters are averaged over 15 wires for signals that are well above the noise. The parameters of EAS events are reconstructed from the data of the surface detectors of the Yakutsk array.

These data should be corrected for the response of the DAS to estimate the dispersion of the input signal. The measured average RMS deviation of signals in 15 wires in the EAS event under consideration is found to be $\sigma_{out}=23.6$ ns. The variance of the output signal is the sum of variances of the input signal and the DAS (\ref{app-c}). Consequently, the RMS deviation of the Cherenkov light signal is estimated as
$$
\sigma_{Cher}=\sqrt{\sigma_{out}^2-\sigma_{DAS}^2}=9\pm5 ~ns.
$$
Note that the time constant of Hamamatsu R2486 PMT, $\tau_{PMT}=5.5$ ns \cite{Hmmts}, is included in the DAS time constant.

\begin{figure}[t]\centering
\includegraphics[width=0.95\columnwidth]{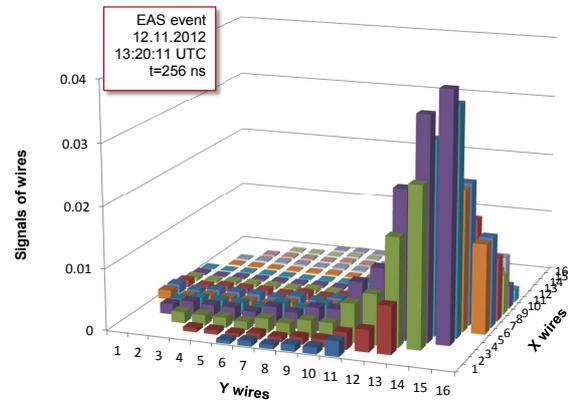}
  \caption{The particular distribution of the Cherenkov signal outputs on the surface of the photo-anode. A grid of wires has a pitch size of 3.75 mm. The distribution is taken at the time $t=256$ ns after the event trigger when the signal amplitudes are approximately maximum.}
  \label{Fig:Tmg}
\end{figure}

The angular distribution of the Cherenkov signal is converted by the telescope into the spatial distribution of the signal on the PMT anode surface. Using the position-sensitive PMT, we are able to measure the spatial distribution of the signal. The spatial resolution of the PMT, $\sigma_R$, is connected with the angular resolution $\sigma_\theta\sim\sigma_R/F$. An example of the spatial distribution of the signal on the anode surface is given in Fig. \ref{Fig:Tmg} at the moment of the maximal amplitude. We used here a particular solution (Eq. 3) considering the distribution of the signal on the anode as symmetric, with a single maximum.

The maximum of the light intensity points to the $x_m^{Cher}$ position in the atmosphere. Combining it with EAS axis coordinates determined by the scintillation counters of the array, we found the height of the maximum to be $h_m\leq1300$ m, corresponding to $x_m^{Cher}\geq850$ g/cm$^2$. In this particular shower, the lower limit to the depth is the result of the maximum of light intensity located at the edge of the telescope aperture.

In general, having a set of WFOV telescopes deployed at the EAS array, one is able to locate the shower maximum based on the angular distribution of the Cherenkov signals detected. Another method of the shower profile reconstruction is evident from the results of Cherenkov pulse shape measurements illustrated in Fig. \ref{Fig:Pulses}. In this case, the position of the maximum in the temporal spread of signals, $t_{max}$, can be connected to the time of maximal radiation of Cherenkov photons. The method will be discussed in the next section.

\section{Modernization of the Cherenkov light detectors of the Yakutsk array}
At present, the Yakutsk array is equipped with 48 Cherenkov light detectors consisting of an FEU-49 PMT mounted vertically in a light-proof container with a motorized lid \cite{NJP}. Recently, 3 differential detectors of the camera obscura type were deployed \cite{Ykt}; the Ethernet LAN of the array connecting all detectors was upgraded: coaxial cables were replaced by optical fibers of 1 Gbps capacity.

Our plan for modernization of the Cherenkov light detectors consists of two steps:
\begin{itemize}
\item to deploy a number of WFOV telescopes in stations calibrating the output signals to that of the existing Cherenkov light detectors in the same stations;
\item to replace the distributed timing and data network hardware and software to synchronize the detectors with an accuracy of nanosecond order compared to the central time server.

\end{itemize}

\begin{figure}[t]\centering
\includegraphics[width=\columnwidth]{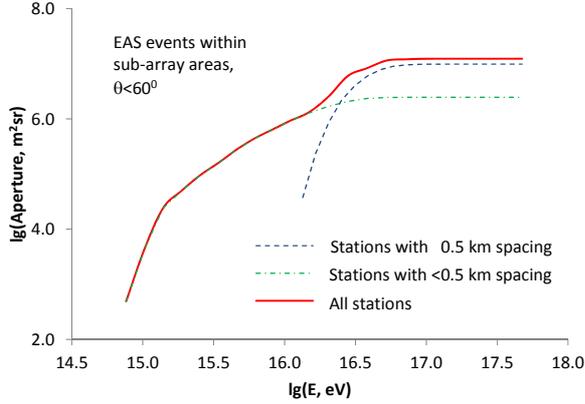}
  \caption{Sub-array aperture of the air Cherenkov light detectors of the Yakutsk array.}
  \label{Fig:Seff}
\end{figure}

With this plan, we have a predefined detector deployment specified by existing stations (Fig. \ref{Fig:Map}). The entire Cherenkov sub-array aperture is determined by the FOV and the aperture of the detector and telescope, in addition to the number of detectors and a set filling. We have calculated the upgraded sub-array aperture (Fig. \ref{Fig:Seff}), assuming the same number of stations as in \cite{NJP}.

The irregular spacing of the stations, being more compact at the center, leads to the effective area of the array increasing with energy. This dependence of the effective area on energy can somewhat compensate for the intensity of CRs decreasing as $E^{-3}$. Above $10^{17}$ eV, the aperture is constant for the showers with axes within the array area.

With a set of telescopes deployed at array stations, an additional improvement of the $x_m^{Cher}$ location accuracy is apparent due to the stereoscopic effect in EAS events detected using two or more telescopes simultaneously. To take advantage of the stereoscopic effect and to increase the detection area, at least 19 telescopes with a spacing of 500 m are required in the Cherenkov sub-array. In this case, above $10^{17}$ eV, the sub-array effective area is $\sim2.6$ km$^2$. The annual number of EAS events detected with 19 telescopes and the array simultaneously is estimated as $\sim3500$ based on the numbers given in Section 4.

\subsection{Reconstruction of $x_m^{Cher}$ based on the Cherenkov light measurements}
Presently, in the Yakutsk array group, the position of the shower maximum in the atmosphere is estimated using a model relation between the lateral distribution slope of Cherenkov light and $x_m$ \cite{NIMA,MSU}. Another method in use here and in the Tunka collaboration is an implementation of the approach described in \cite{Fmn} -- a model-independent relationship between the Cherenkov signal FWHM and $x_m$.

After enhancement of the Yakutsk array with a set of WFOV Cherenkov telescopes, we will be able to use a new method -- direct calculation of the $x_m^{Cher}$ position in the atmosphere based on the Cherenkov signal delay timing in stations, besides the geometrical triangulation of $x_m^{Cher}$ via the shower light profile mirrored in telescopes. The method can be used concurrently with or instead of the methods mentioned above. A distinctive feature of the new method is a combination of the pulse shape measurement with the time delay of Cherenkov signals in detectors.

Indeed, having the time difference (delay time), $\Delta t$, between distances from $x_m^{Cher}$ to a detector and to the shower core on the array plane, one can calculate the height of the maximum, $h_m$, by solving the equation:
$$
c\Delta t=\sqrt{h_m^2+(h_m\tan\theta+R\cos\alpha)^2+R^2\sin^2\alpha}-h_m\sec\theta,
$$
where $R$ is the EAS core distance; the air refraction index is assumed as 1 (the approximation is given in \ref{app-b}).

\begin{figure}[t]\centering
\includegraphics[width=\columnwidth]{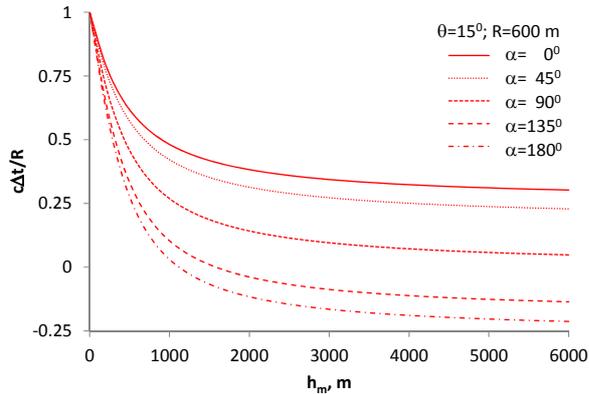}
  \caption{Time difference of the Cherenkov photon arriving from $x_m^{Cher}$ to the detector and to the shower core in the array plane.}
  \label{Fig:cDt}
\end{figure}

The equation has a solution given by
$$
h_m=0.5\frac{R^2-(c\Delta t)^2}{c\Delta t\sec\theta-R\tan\theta\cos\alpha}.
$$
For illustration, a family of curves $c\Delta t/R(h_m)$ is shown in Fig. \ref{Fig:cDt} with different angles between the detector and the shower axis projection on the array plane. The time difference is bounded within the allowed values from the smallest at $h_m\rightarrow\infty$ to the largest $R/c$ at $h_m=0$.

The delay time of Cherenkov signals at the maximum can be measured by the set of WFOV telescopes. As was demonstrated in the previous section, the Cherenkov pulse shape is sampled by a number of independent telescope channels in the individual EAS events. This provides the accurate location of the signal maximum and delay timing.

Secondly, a synchronization of telescopes with nanosecond accuracy is required in addition to the accurate pulse shape measurement of Cherenkov signals, because of the minimal value of the pulse width near the shower core (Fig. \ref{Fig:PulseWidth}). A set of synchronized telescopes can be used to locate the shower axis position in the array plane using either the Cherenkov light intensity or the FWHM of the signal as a function of the axis distance \cite{Tunka3}.

The accuracy of $x_m^{Cher}$ determination depends on the accuracies of the delay timing and core distance measurements. We estimated the accuracy assuming $\alpha=0$ and $\theta=0$ in the isothermal atmosphere, for simplicity. We also used the relative contributions of the depth intervals in the atmosphere to the Cherenkov light flux detected at the core distance $R$ calculated in Section 2 (illustrated in Fig. \ref{Fig:Xcontribution}). The results are presented in Table \ref{Table:3} for the optimistic and pessimistic estimations of RMS errors in the measurements of time delay and distance. The accuracy of $x_m^{Cher}$ estimation gets better with increasing shower axis distance due to decreasing relative errors in $\Delta t$ and $R$.

The main advantage of the direct $x_m^{Cher}$ calculation method is model-independence. The method is applicable to all Cherenkov light detectors having sufficient temporal resolution, while the geometrical triangulation of $x_m^{Cher}$ is possible with telescopes only.

\subsection{Estimation of the mass composition of cosmic rays}
A pulse shape of Cherenkov light from an EAS measured with a detector at the particular shower core distance $R$ is formed mainly by the shower development profile and the angular distribution of electrons. The former part is model-dependent, typically parameterized by the maximum of the shower profile, $x_m$. The latter can be calculated in electromagnetic cascade theory and is independent of nuclear interaction characteristics and the mass composition of CRs.

Measurements of the Cherenkov signal shape at distances $R>600$ m from the shower core can be used to locate the $x_m$ position in the atmosphere. However, the more immediate approach is to focus on $x_m^{Cher}$ observable at all the possible distances from the core. Because of the universal angular distribution of electrons, the Cherenkov signal maximum is a function of the EAS longitudinal profile, which is sensitive to the primary particle mass.

In this way, for example, one can determine the energy dependence of the average CR mass composition measuring $x_m^{Cher}$ in different energy intervals.

Essentially, the $x_m^{Cher}$ correlation to the shower $x_m$ is model-dependent \cite{Ztspn,Trvr}. On the other hand, the shower age and the mass composition of EAS primaries are inherent parameters of the Cherenkov light pulse shape. So, the preferred characteristic of the shower profile observed with Cherenkov light detectors is $x_m^{Cher}$.

\section{Summary}

We designed and assembled the engineering prototype of the WFOV Cherenkov telescope to operate in cooperation with the surface detectors of the Yakutsk array. Field testing of the telescope demonstrated the practical applicability of the ideas proposed; the performance agrees with the main expectations from simulations.

During the winter of 2013-2014, a number of EAS events were detected by the telescope in coincidence with the surface detectors subset of the Yakutsk array. The detection efficiency of the telescope was measured, as well as the effective radius of the telescope detecting area. The angular and temporal structural parameters of the air Cherenkov light signal from the particular EAS events were measured.

A way of modernizing the Yakutsk array is proposed by adding a set of WFOV telescopes. In conjunction with the fast synchronization system of the array detectors, it will provide quite new possibilities in measuring the mass composition of cosmic rays in the energy range above $10^{15}$ eV.

\begin{table}[t]\centering
\caption{Estimation of the $x_m^{Cher}$ reconstruction accuracy, g/cm$^2$.
$E_0=10^{15}$ eV.}
\begin{tabular}{rrrr}
\hline
          $R$, m  &  600 &  400 &   200 \\
 $\sigma_t= 3$ ns &  5.8 &  9.1 &  12.3 \\
 $\sigma_t=10$ ns & 19.4 & 30.2 &  41.0 \\
 $\sigma_R= 5$ m  &  5.3 &  7.0 &   9.0 \\
 $\sigma_R=50$ m  & 53.0 & 70.0 &  89.9 \\
\hline
\end{tabular}
\label{Table:3}\end{table}

\section*{Acknowledgments}
We are grateful to the Yakutsk array staff for the data acquisition and analysis. We would like to thank the anonymous referee(s) for the critical comments and hints for improving the English. The work is supported in part by the RAS (program 10.2) and RFBR (grants 11-02-00158, 12-02-31550, 13-02-12036).

\appendix
\section{An algorithm of Cherenkov signal modeling in EAS}
\label{app-a}

We used the assumptions of V.I. Zatsepin \cite{Ztspn} combined with conclusions of the Haverah Park group \cite{Trvr} to model the Cherenkov light emission by relativistic electrons in the atmosphere ($R_i>$ 200 m): the energy spectrum of electrons is taken at the shower maximum; the spatial distribution of the electrons is neglected; the angular distribution of the electrons is Gaussian; the difference in the time of flight for photons to the detector and for the shower axis to the array plane is given by $ct=L_d-L$, where $L$ is the distance from a point of emission to the array along the shower axis; $L_d$ is the distance to the detector; $c=0.3$ m/ns; the influence of the refraction coefficient in air is neglected. In \ref{app-b}, the applicability domain of the approximation $n=1$ is bounded.

For a detector with the impact parameter equal to the distance to the shower axis, $R_i$, the distance from the emission point to the impact point\footnote{the closest point to detector in the shower axis} is:
$$
L_i=L-R_i\cos\alpha\sin\theta/\sqrt{1-\cos^2\alpha\sin^2\theta},
$$
where $\theta$ is the zenith angle; $\alpha$ is the azimuthal angle between the detector and the shower axis projection on the array plane. The photon number in the detector is
$$
\frac{dQ_d}{dx}\propto a(x-x_0)\frac{S_dL\cos\theta}{L_d^3}\int_{E_{th}}^{E_0}dE N(x,E,\nu)(1-\frac{E_{th}^2}{E^2}),
$$
where $a(x-x_0)$ is the absorption coefficient of light in the atmosphere between $x$ and the array level $x_0$; $S_d$ is the detector area; $N(x,E,\nu)$ is the distribution of electrons in energy and angle; $\nu=\arctan\frac{R_i}{L_i}$; $E_{th}\approx21.1\sqrt{x_0/x}$, MeV is the threshold energy of Cherenkov radiation \cite{JETP}; $E_0$ is the primary particle energy.

In the program, the last formula is coded.

\section{Influence of the refractive index in the atmosphere}
\label{app-b}

In a vertical shower, the arrival time of photons at the detector with respect to the shower axis crossing the array plane is given by: $tc+h=\frac{\sqrt{h^2+R^2}}{h}\int_0^hn(q)dq$, where $h$ is the height of a shining point; $R$ is the shower axis distance. In an inclined shower, the situation does not differ significantly from that of a vertical shower. According to the Gladstone-Dale relation, $n=1+(n_0-1)\rho/\rho_0=1+(n_0-1)exp(-h/h_0)$, where $n_0, \rho_0$ are the refractive index and the air density at the array level, respectively. Integrating the equation, one has $tc+h=\sqrt{h^2+R^2}(1+(n_0-1)\frac{h_0}{h}(1-exp(-h/h_0)))$.

Thus, the influence of the air refractive index $n\neq 1$ is estimated as $(n_0-1)(1-0.5h/h_0)\leq 3\times 10^{-4}$, where $h<h_0=6.9$ km.

\section{Properties of the detector response function}
\label{app-c}
When the detector input signal is a $\delta$-function located at $t_i$, the DAS output signal is a function $g(t-t_i)$ with some time constant. An arbitrary input function can be represented as a sum of $\delta$-functions:
$$
f_{in}(t)=\lim_{N\rightarrow\infty}\sum_{i=1}^N f_{in}(t_i)\delta(t-t_i)=\int_{-\infty}^{\infty} f_{in}(\tau)\delta(t-\tau)d\tau.
$$
The detector transforms every $\delta$-function into $g(t-t_i)$
$$
f_{out}(t)=\lim_{N\rightarrow\infty}\sum_{i=1}^N f_{in}(t_i)g(t-t_i)=\int_{-\infty}^{\infty} f_{in}(\tau)g(t-\tau)d\tau.
$$
Hence, mathematically, our detector is equivalent to the linear integral operator with a difference kernel function that transforms $f_{in}$ to $f_{out}$.

The main properties of the convolution operator of interest to us are that the mean values and the variances of the functions are additive:
$$
\overline{f_{out}}=\overline{f_{in}}+\overline{g}
$$
$$
\sigma^2_{f_{out}}=\sigma^2_{f_{in}}+\sigma^2_{g}.
$$

\end{document}